# Microfluidics for Chemical Synthesis: Flow Chemistry

*Klavs F. Jensen - Massachusetts Institute of Technology*

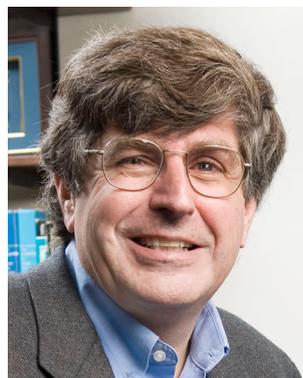

*Klavs F. Jensen is Warren K. Lewis Professor in Chemical Engineering and Materials Science and Engineering at the Massachusetts Institute of Technology. From 2007- July 2015 he was the Head of the Department of Chemical Engineering. He received his Ph.D. in chemical engineering from the University of Wisconsin-Madison. His research interests revolve around reaction and separation techniques for on-demand multistep synthesis, as well as microsystems biological discovery and manipulation. Catalysis, chemical kinetics and transport phenomena are also topics of interest along with development of simulation approaches for reactive chemical and biological systems. He is the co-author of 400 refereed journal and 175 conference publications and 45 US patents. He chairs the Editorial Board for the new Royal Society of Chemistry Journal Reaction Chemistry and Engineering. He serves on advisory boards to universities, companies, professional societies, and governments. He is the recipient of several awards, including a National Science Foundation Presidential Young Investigator Award, a Camille and Henry Dreyfus Foundation Teacher-Scholar Grant, a Guggenheim Fellowship, and the Allan P. Colburn, Charles C.M. Stine, R.H. Wilhelm, W.H. Walker, and Founders Awards of the American Institute of Chemical Engineers. He received the inaugural IUPAC-ThalesNano Prize in Flow Chemistry in 2012. Professor Jensen is a member of the US National Academies of Sciences and Engineering and the American Academy of Arts and Science. He is a Fellow of the American Association for the Advancement of Science (AAAS), the American Chemical Society, and the American Institute of Chemical Engineers, and the Royal Society of Chemistry.*

## Abstract

Chemical synthesis in microfluidic systems has matured over the two past decades from simple demonstration examples to applications in pharmaceuticals and fine chemicals.[1] Advantages of controlled mixing, enhanced heat and mass transfer, expanded reaction conditions, and safety have driven adoption of continuous flow techniques. The field has moved beyond single transformations to continuous multistep synthesis of active pharmaceutical ingredients by incorporating in-line workup techniques (Figure 1).[2] Moreover, integration of on-line measurements of reactant flows, reactor temperature, and outlet concentrations with feedback control systems has enabled automated optimization of reaction yields as well as determining kinetic information.[3]

The evolution of microfluidic systems is reviewed with focus on application for which microfluidics offers particular advantages in terms of increased heat and mass transfer, high

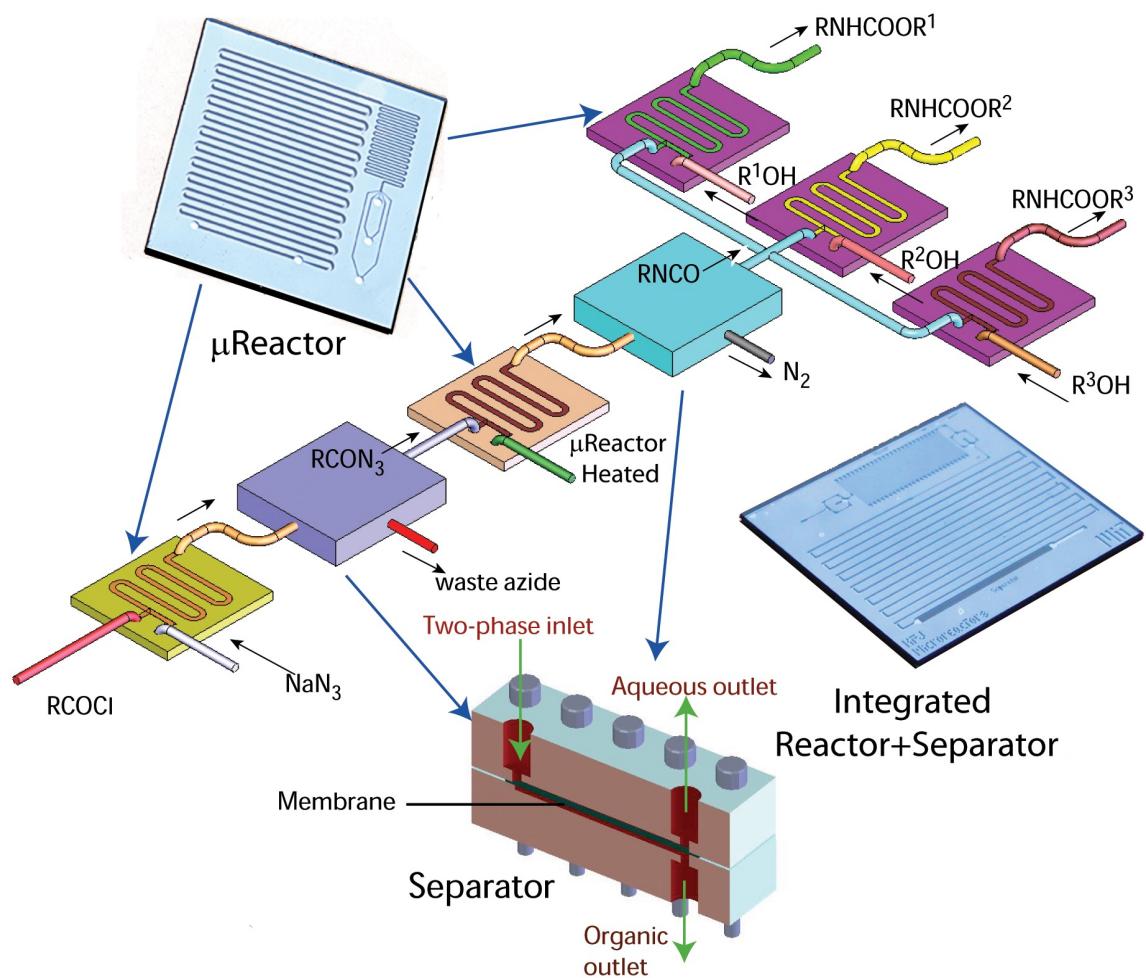

*Figure 1. Multi-step microfluidic chemical synthesis of carbamates starting from aqueous azide and organic acid chloride using the Curtius rearrangement reaction. The scheme involves three reaction steps and two separation steps.[2]*

temperature and high pressure operation, and synthesis involving highly reactive and toxic intermediates.[4] In addition to applications in fine chemicals and pharmaceuticals, microfluidic systems have become tools for synthesis of nanoparticles by allowing reproducible synthesis and access to conditions not easily accessed in batch, e.g., supercritical conditions.[5] The systems also serve as effect methods to understand the kinetics and early stage growth of nanoparticles, including quantum dots. Application examples include microfluidic control of metal nanoparticle shape and synthesis of InP Quantum dots.

As an example of current efforts, an automated droplet microfluidic screening and optimization technique (Figure 2) enables precise control of reaction temperature and residence time for each experiment run in microliter-scale droplets. The droplets are prepared by a computer-controlled liquid handler to enable screening of discrete variables (e.g., catalyst species, base, solvents) without requiring system depressurization or reconfiguration.[6] The automation and control strategy further enables high reproducibility and flexible residence time with similar mixing and mass transfer characteristics. Coupling the droplet reactor with a light source converts the system into an automated microfluidic platform for in-flow studies of visible-light photoredox catalysis.[7] The micro fluidic droplet platforms is demonstrated through a number of case studies, including catalyst optimization, N-C bond formation, photocatalytic oxidation, and synthesis of the drug substances.

Scaling of chemical synthesis in microfluidics enables a plug-and-play, reconfigurable, refrigerator-sized manufacturing platform for on-demand synthesis of common pharmaceuticals.[8] In this flexible system multi-step synthesis occurs at elevated temperatures and pressures to enhance reaction rates, and the resulting residence times are a few minutes, in contrast to the multiple hour-long synthesis typically needed for batch. Typical production rates are grams/hour sufficient to produce thousands of doses per day of common pharmaceuticals. Lastly, future opportunities are integration of machine learning for synthesis planning[9] with fully automated, sel-configuring microfluidic chemical synthesis systems.

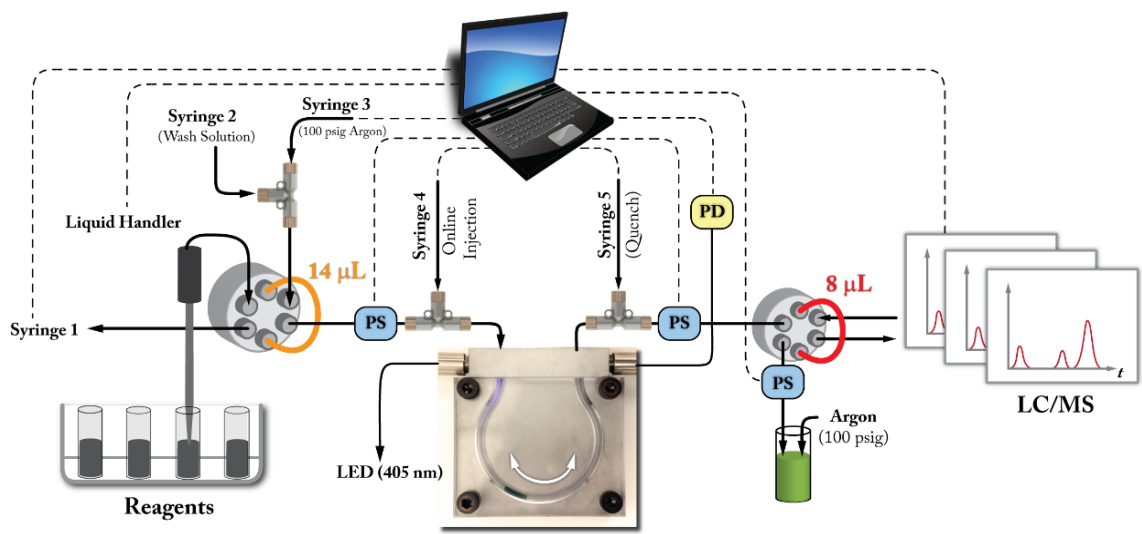

*Figure 2. Schematic of the automated droplet-based reaction screening platform. PS: phase sensor; PD: photodetector.[7]*